\documentclass[a4paper,fleqn]{cas-sc}
\usepackage[numbers]{natbib}
\def\tsc#1{\csdef{#1}{\textsc{\lowercase{#1}}\xspace}}
\tsc{WGM}
\tsc{QE}
\tsc{EP}
\tsc{PMS}
\tsc{BEC}
\tsc{DE}
\usepackage{stackrel}

\newcommand{\sugb}[1]{\textcolor{blue}{#1}}

\begin{document}
\let\WriteBookmarks\relax
\def\floatpagepagefraction{1}
\def\textpagefraction{.001}
\shorttitle{Physical-chemical Approach for the Impact of Modifying Molecular Bridges of \dots}
\shortauthors{Madrid et~al.}
\title [mode = title]{Physical-chemical Approach for the Impact of Modifying Molecular Bridges of TPA-Based Systems to Improve the Photovoltaic Properties of Organic Solar Cells}                      

\author[1]{D. Madrid-Úsuga}[
orcid=https://orcid.org/0000-0002-4916-0211]
\cormark[1]
\ead{duvalier.madrid@unisucrevirtual.edu.co}
\credit{Conceptualization, Numerical simulations and writing of the manuscript}
\address[1]{Departamento de F\'{i}sica, Universidad de Sucre, A.A. 406, Sincelejo, Colombia.\\ Grupo de Investigación Teoría de la Materia Condensada}

\author[1]{ O. J. Suárez}
\credit{Conceptualization, and writing of the manuscript}


\begin{abstract}
The theoretical design of donor chromophores based on triphenylamine and 2-(1,1-dicyanomethy lene)rhodanine (\textbf{DCRD-DCRD-2}) is proposed through structural adaptation with several molecular bridges derived from thiophene that can be used as new organic materials for organic solar cells (OSC). The optoelectronics properties and geometries of the \textbf{DCRD-DCRD-2} organic molecules are characterized using the B3LYP and CAM-B3LYP functional, with the basis set 6-31G(d,p). Consequently, the UV-Visible results revealed that a good relationship was found between the experimental values and the calculated using the DFT and TD-DFT level of theory. The study involved the prediction of photo-physical descriptors such as frontier molecular orbitals, ionization potential, electron affinity, molecular electrostatic potential, reorganization energy, open circuit voltage ($V_{oc}$), fill factor (FF), and short-circuit current ($J_{sc}$) in the ground state geometry, using the B3LYP/6-31G(d,p) basis set. Structural tailoring with various molecular bridges resulted in a narrowing of the energy gap ($2.130$ -- $1.96$~eV) with broader absorption spectra ($525.55$ -- $417.69$~nm). An effective charge transfer toward the lowest unoccupied molecular orbitals (LUMO) from the highest occupied molecular orbitals (HOMO) was studied, which played a crucial role in conducting materials. \textbf{DCRD-2} exhibited $\lambda_{max}$ at $417.69$~nm in EtOH (ethanol) solvent with the lowest band gap ($1.96$~eV) and the lowest excitation energy of $2.968$~eV. The highest mobility of holes and electrons is determined in all the designed molecules due to their low reorganization energy values that validated preferable photovoltaic properties in the \textbf{DCRD-1} molecular system. \textbf{DCRD-1} showed the lowest electron mobility ($\lambda_{e} = 0.317$~eV) and \textbf{DCRD-2} demonstrated the lowest hole mobility ($\lambda_{h} = 0.198$~eV), it is found that the \textbf{DCRD-1} molecular system presents better equilibrium properties for the transport of holes and electrons. Our comparative results are proposed as possible options for designing efficient OSCs. Finally, it was found that by using the molecular systems in the active layer of a bulk heterojunction (BHJ) organic solar cells as electron-donor materials, the \textbf{DCRD-1} compound shows more efficient results for $V_{oc}$, $J_{sc}$ and FF and, therefore, it has a slightly higher energy conversion power compared to \textbf{DCRD} and \textbf{DCRD-2} compounds.\\
\end{abstract}

\begin{graphicalabstract}
\includegraphics[keepaspectratio, width = 80mm]{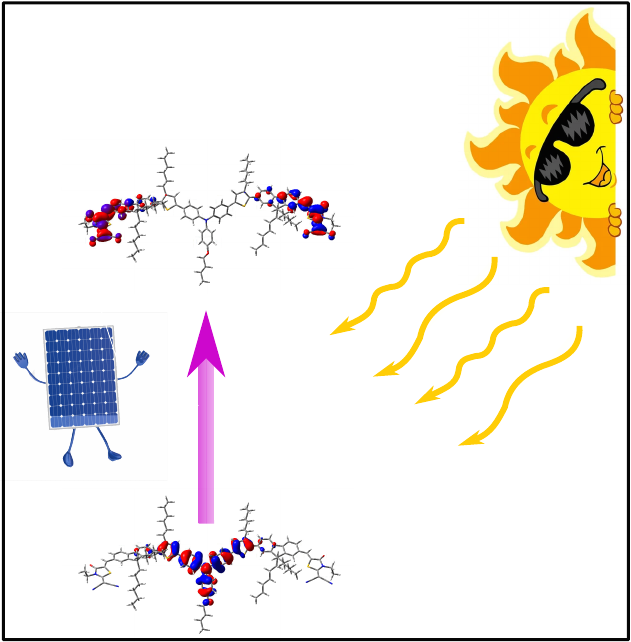}
\end{graphicalabstract}

\begin{highlights}
\item Triphenylamine derivatives were characterized using density functional theory (DFT) and time-dependent (TD-DFT) calculations.
\item We have studied four A-$\pi$-D-$\pi$-A molecular system that contain triphenylamine, thiazole and thiophene fragment in their structure.
\item The estimation of the energy of the HOMO-LUMO boundary molecular orbitals showed an important correlation with the available experimental data.
\item The calculated molecular systems present an important potential for use as photovoltaic materials in BHJ solar cell.
\end{highlights}


\maketitle
\section{Introduction}

Currently, the excessive exploitation of natural resources, the growth of global energy demand, and the environmental crisis, associated with the excessive increase in fossil fuels, generate concern for immediate attention and show the need for alternative sources of renewable energy to stop further impacts. In this sense, organic solar cells (OSCs) are potential candidates to help in this task, due to the large amount of sunlight available. Silicon with high thermal stability, low toxicity and high efficiency of about 47.6\% has been used as a semiconductor material in solar cells for decades~\sugb{\cite{schygulla2022}}. However, there are certain drawbacks associated with silicon-based solar cells, including non-tunable energy levels, hardness, and high cost with manufacturing difficulty. Therefore, OSCs have been the subject of intense research due to their potential for the generation of renewable energy~\sugb{\cite{ma2020, zheng2022, yuan2022}}. These photovoltaic cells, mainly composed of organic materials, have demonstrated a good ability to convert sunlight into electricity efficiently~\sugb{\cite{liu2019, zhu2022, jia2023}}. Despite their advantages in flexibility~\sugb{\cite{fukuda2020}}, lightness and potential low production cost~\sugb{\cite{yang2021}}, CSOs still face significant challenges in terms of energy efficiency. In the quest to improve the efficiency of OSCs, the creation and design of new organic materials play a fundamental role~\sugb{\cite{hussain2021}}. The design and synthesis of new molecular compounds seek to overcome current limitations by optimizing photo-physical and electrochemical properties, thus improving the absorption of sunlight, the mobility of electrons and the reduction of energy losses~\sugb{\cite{nuhash2020}}. The design of new molecular materials for organic solar cells involves exploring a wide range of chemical structures and configurations to maximize light absorption, improve electron mobility and reduce energy losses. In this sense, triphenylamine (TPA) is a good candidate for this purpose~\sugb{\cite{nhari2021, joseph2022}}, since the use of flat or bulky groups is crucial for the aggregation of the system, which disfavors the charge recombination processes and, therefore, increases the efficiency of the device; which together with push-pull systems covalently linked to 2-(1,1-dicyanomethylene)rhodanine~\sugb{\cite{echeverry2014}}, allows efficient anchoring to TiO, since it prevents dissociation on the TiO surface after long irradiation time, increasing the stability of the molecules~\sugb{\cite{tigreros2021}}.
\medskip

In this paper, we have designed three new A-$\pi$-D-$\pi$-A type chromophores, namely \textbf{DCRD-DCRD-2}, derived from rhodanine. All of the new compounds have been designed by changing the molecular bridging unit of the chromophores, which contain two (1,1-dicyanomethylene)rhodanine units at the terminal part linked at the terminal with an electron-donating nucleus, triphenylamine. We studied the impact of various molecular bridging units on the electronic, optical, and charge transfer behavior of the newly designed chromophores to evaluate them as efficient OSC-based composites for future manufacturing of high-performance electronic devices.
\medskip

This article is organized as follows: In section \sugb{\ref{system}}, se muestran los sistemas moleculares diseñados para este estudio y el metodo computacional implementado, section~\sugb{\ref{results}}, the results associated with the study systems are presented, such as: Frontier molecular orbital (FMO); Absorption spectral, ionization potential and electron affinity, reorganization energy and molecular electrostatic potential, in section~\sugb{\ref{photovoltaic}} the photovoltaic properties as well as the energy conversion capacity (PCE) parameter of the study systems are presented. as hole transporting material. Finally, in section~\sugb{\ref{Conclusions}}, we present our conclusions.

\section{Chromophores and Computational Methods}\label{system}

\begin{figure}[htp]
\centering
\includegraphics[scale=0.32]{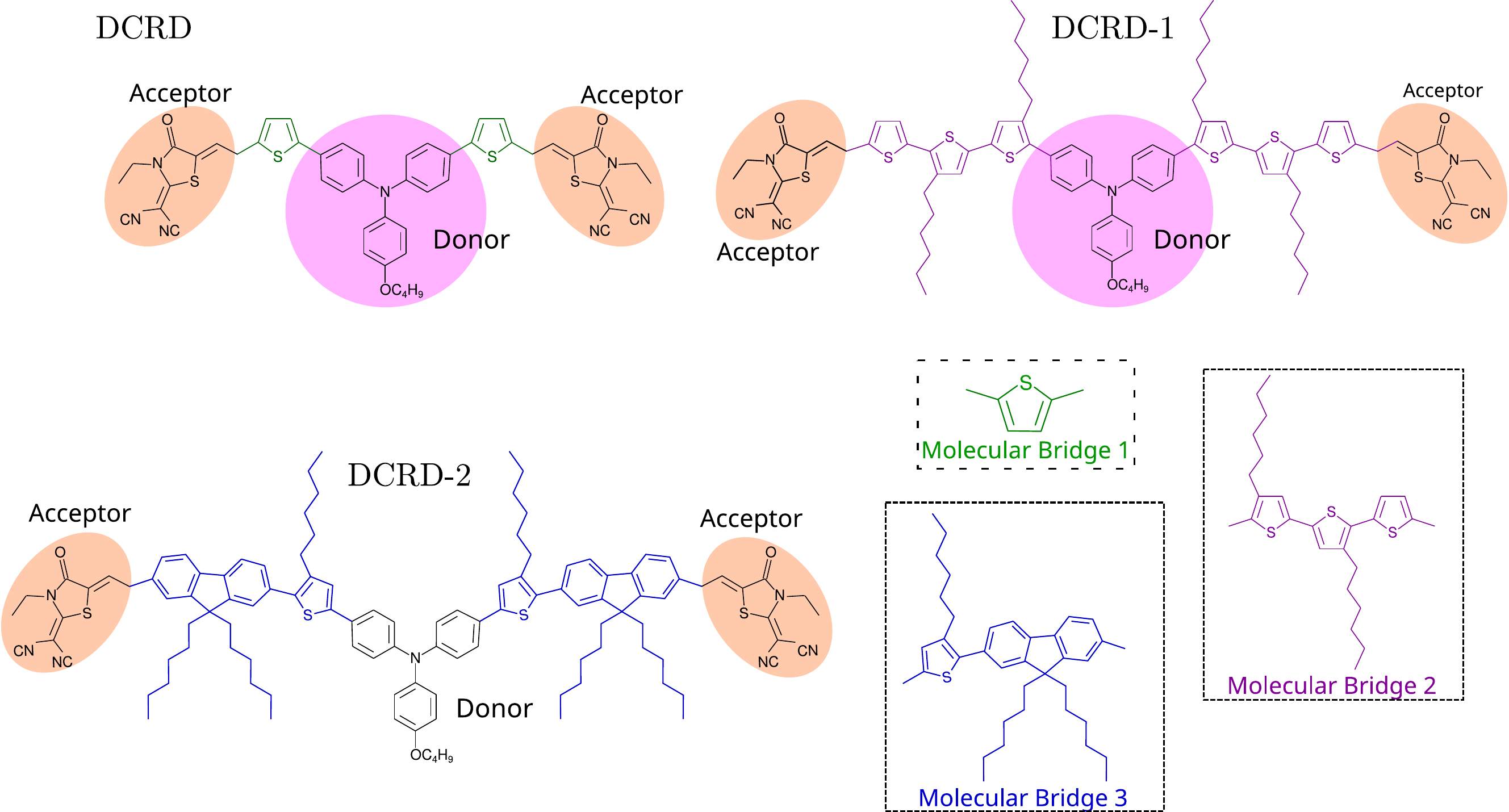}
\caption{Molecular Chromophores under study. For simplicity reasons we labeled the complexes as \textbf{DCRD}: made up of a fragment of 4--(Hexyloxy)triphenylamine (Donor) + 2,5--Dimethylthiophene (Bridge 1) +2--(1,1-Dicyanomethylene)--1,3--thiazolene--4 (Acceptor); the \textbf{DCRD-1}: is made up of 4--(Hexyloxy)triphenylamine (Donor) + 4,4''--Dihexyl--2,2':5',2''--terthiophene (Bridge 2) + 2--(1,1-Dicyanomethylene)--1,3--thiazolene--4 (Acceptor); and the \textbf{DCRD-2}: is 4--(Hexyloxy)triphenylamine (Donor) + 2--(9,9-Dihexil--9H--fluoren--2--yL)hexylthiophene (Bridge 3) + 2--(1,1--Dicyanomethylene)--1,3--thiazolene--4 (Acceptor).}
\label{Fig_1}
\end{figure}
In this work, we consider an organic molecular system as presented in Fig.~\sugb{\ref{Fig_1}}. They classsify each molecule into three fragments, that is to say, a structure that acts as an electron-donor (central unit), molecular bridges to facilitate intermolecular charge transfer, followed by electron-acceptors (terminal groups). In the systems, we use the structure of 4-Hexylocy-Triphenylamine (\textbf{TPA}) as an electron-donor fragment consisting of a Triphenylamine nucleus, connected by a structure of molecular bridges (\textbf{B}) to the fragment electron-acceptor 2-(1,1-Dicyanomethylene)-1,3-thiazolene-4 (\textbf{DCR}). We use different molecular bridges for the final design of the new push-pull chromophores, such as 2,5-Dimethylthiophene (\textbf{B1}), 4,4''-Dihexyl-2,2':5',2''-Terthiophene (\textbf{B2}) and 2-(9,9-Dihexyl-9H-Fluoren-2-yl)hexylthiophene (\textbf{B3}); to design the systems, (\textbf{DCRD}), (\textbf{DCRD-1}) and (\textbf{DCRD-2}), respectively. To obtain the geometry of the most stable conformation of the molecular systems, they were geometrically optimized to give an image of the molecular structures in the ground state ($S_0$) using density functional theory (DFT) through the hybrid functional B3LYP~\sugb{\cite{civalleri2008}}, with the basis set 6-31+G(d,p), using Gaussian 09~\sugb{\cite{frisch2016}}. The energies of the molecular system are calculated with the optimized structures. The excited states of the molecules, as well as the calculation of the UV-Vis spectra reported here, were simulated using of time-dependent density functional theory (TD-DFT) and the CAM-B3LYP/6-31+G(d,p) basis set. We considered the systems in the gas phase and the presence of an ethanol environment (EtOH) as a solvent by means of the conductor-like polarization continuum model (C-PCM)~\sugb{\cite{takano2005, mennucci2012}}. It was found that the functional showed a reasonable affinity with the experimentally reported maximum absorption wavelength $\lambda_{max}$ for reference \textbf{DCRD} systems. Additionally, we derive different properties of molecular systems, such as higher occupied molecular orbitals (HOMOs), lower unoccupied molecular orbitals (LUMOs), energy gap, reorganization energy, and excitation energy, from the computational results.

\section{Results and Discussion}\label{results}

\subsection{Frontier Molecular Orbital (FMO) Analysis.}
\medskip

\begin{figure}[t]
\centering
\includegraphics[scale=0.4]{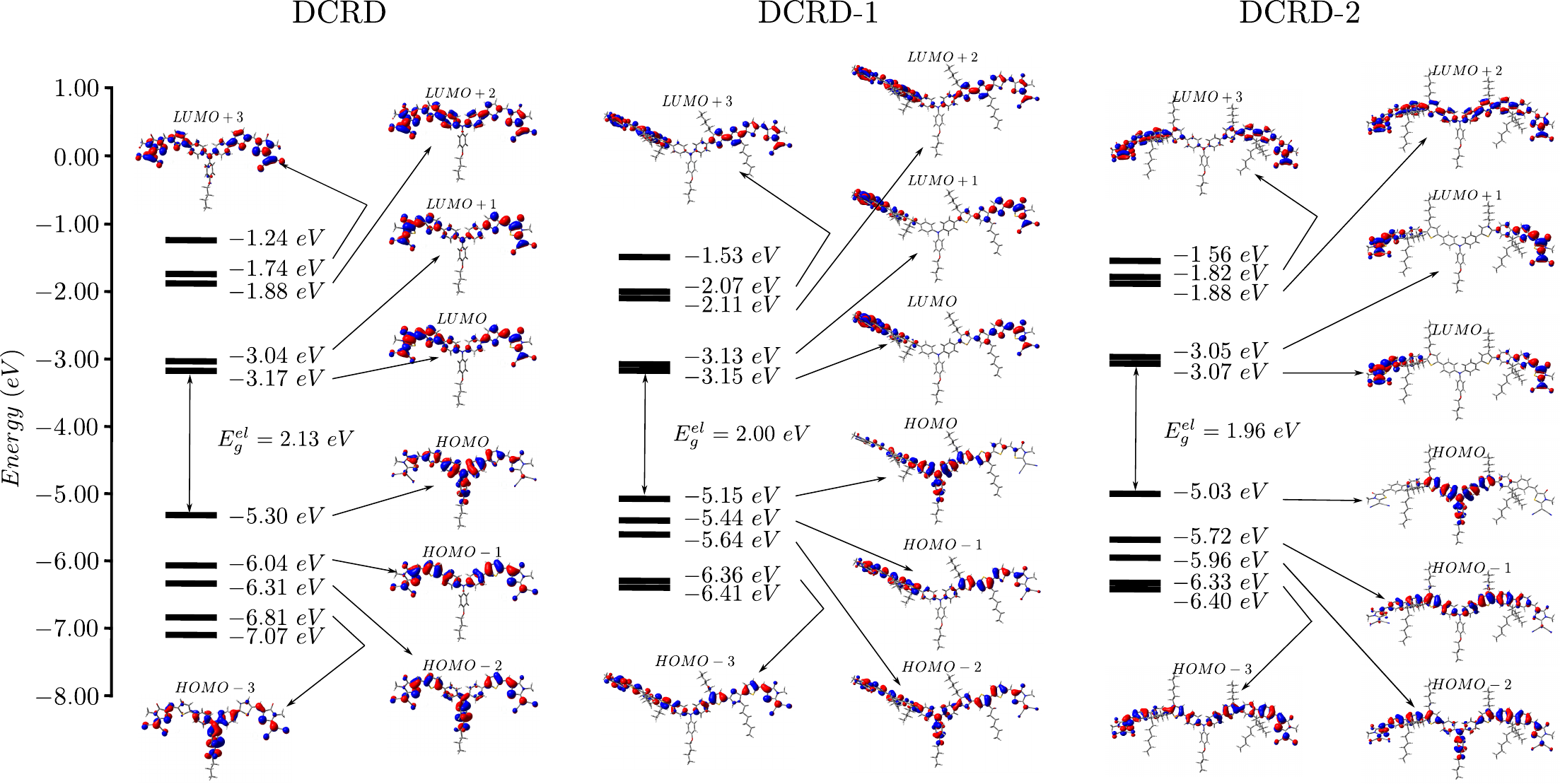}
\caption{Energy level diagram of the Kohn-Sham orbitals of the dyes \textbf{DCRD}, \textbf{DCRD-1}, \textbf{DCRD-2} calculated using density functional theory (DFT) for the EtOH solvent.}
\label{Fig_2}
\end{figure}

Starting from the optimized geometries of the molecular systems \textbf{DCRD}, \textbf{DCRD-1} and \textbf{DCRD-2}, we implemented TD-DFT to determine the electronic transitions that best favor charge transfer processes (CT) in molecular systems, as well as analyzing the FMO frontier molecular orbitals (i.e., HOMOs and LUMOs) in order to quantify the relationship between structural and electronic geometry, FMOs are crucial variables in quantum physics and chemistry where they are used to characterize charge transfer, molecular correlation, kinetic stability, and chemical reactivity~\sugb{\cite{janjua2012}}. The distribution pattern of FMOs provides valuable images and numbers for characterizing the photovoltaic and optoelectronics behavior of chromophores in OSCs~\sugb{\cite{zhang2020, zahid2021}}. The energy gap ($E_g^{el}$) of FMO indicates the requirement for exciton dissociation; therefore, it is a characteristic quantity of solar cells and other photovoltaic devices~\sugb{\cite{ans2018}}. Molecular systems with broader $E_g^{el}$ are considered more stable, less reactive and harder molecules, while molecules with reduced $E_g^{el}$ are more reactive, softer molecules and that have impressive heat transfer. Intermolecular charge (TIC)~\sugb{\cite{li2014}}. To examine the impact of molecular bridges on the optoelectronics properties of the investigated chromophores \textbf{DCRD}, \textbf{DCRD-1} and \textbf{DCRD-2}, the energy levels of HOMO, LUMO and their band gaps $E_g^{el}$ are analyzed. Therefore, the energies and contour diagrams of the selected Kohn-Sham molecular orbitals, including HOMO and LUMO, are illustrated in Fig.~\sugb{\ref{Fig_2}} for the reference \textbf{DCRD} and \textbf{DCRD-1}, \textbf{DCRD-2} in the environment of an ethanol solvent.
\medskip

For the HOMO in all systems, one can observe that the electrons are significantly localized in the electron-donor fragment of triphenylamine both in the gas phase and in the ethanol solvent (Fig. ~\sugb{\ref{Fig_2}}), while HOMO-1 is mainly composed of fragments of the molecular bridge and the electron-acceptor fragment for the reference system \textbf{DCRD}, and for \textbf{DCRD-1} and \textbf{DCRD-2} focuses mainly on the molecular bridge and a weak contribution from the electron-acceptor fragment, both in the gas phase and in EtOH. However, in the case of the LUMO and LUMO+1 orbitals, these are formed mainly through the participation of the electron-acceptor unit. Therefore, the transitions of molecular systems from HOMO to LUMO, together with the transitions from HOMO to LUMO+1, represent charge transfers from the electron-donor fragment to the electron-acceptor fragment; however, transitions from HOMO-1 to LUMO or LUMO+1 represent CT between the molecular bridge and the electron-acceptor fragment. The HOMO and LUMO energy values of the reference \textbf{DCRD} in EtOH are $-5.30$~eV and $-3.17$~eV, which are in agreement with the experimental reported values ($-5.40$~eV and $ -3.11$~eV)~\sugb{\cite{echeverry2014}}, and also present a band gap of $2.13$~eV which is considered the highest $E_g^{el}$ among all molecular architectures. The estimated HOMO and LUMO energy values for \textbf{DCRD-1} and \textbf{DCRD-2} molecules are shown in Table~\sugb{\ref{tab1}}.
\medskip

\begin{table}[htp] 
\centering\caption{Energies of the HOMO ($E_H$) and LUMO ($E_L$) calculated, maximum absorption wavelength $\lambda_{max}$ and Band Gap ($E_g^{el}$) of all researched molecular systems in Gas Phase and EtOH.}
\scalebox{1.02}{
\begin{tabular}{w{c}{1.6cm}w{c}{1.6cm}w{c}{1.4cm}w{c}{1.4cm}w{c}{1.4cm}w{c}{1.4cm}} 
\hline
\toprule
\normalsize{System} & $\lambda_{max}$ (nm) & $E_{H}$ (eV) & $E_{L}$ (eV) & $E_g^{el}$ (eV) \\ 
\midrule
                 & & \textit{\normalsize{Gas Phase}} & & \\ 
\textbf{DCRD~~~} & 462.40 & -5.58 & -3.28 & 2.30  \\ 
\textbf{DCRD-1}  & 494.83 & -5.28 & -3.18 & 2.10  \\
\textbf{DCRD-2}  & 407.60 & -5.07 & -3.10 & 1.97  \\ 
\midrule
                 &  & \textit{\normalsize{EtOH}}     & & \\ 
\textbf{DCRD~~~} & 449.62 & -5.30 & -3.17 & 2.13  \\ 
\textbf{DCRD-1}  & 525.55 & -5.12 & -3.15 & 1.97  \\ 
\textbf{DCRD-2}  & 417.69 & -5.01 & -3.05 & 1.96  \\
\bottomrule
\hline
\end{tabular}}
\label{tab1}
\end{table}

On the other hand, it is observed that \textbf{DCRD-2} exhibits a narrow band gap of $1.96$~eV in EtOH and $1.97$~eV in the gas phase due to the molecular bridge of 2-(9,9-Dihexyl-9H-Fluoren-2-yl)hexylthiophene and increase in the $\pi$-conjugation system that shows the maximum charge transfer from the HOMO to the LUMO, being the lowest among the other systems that present $E_g^{el}$ of $2.13$~eV for the reference system \textbf{DCRD} and $1.97$~eV for \textbf{DCRD-1}, respectively. Generally, molecules possessing a electron-withdrawing group have a reduced band gap and higher absorption in the UV-Vis region, thus promoting charge transfer~\sugb{\cite{haroon2021}}. The decrease in the $E_g^{el}$ of the dyes \textbf{DCRD-1} and \textbf{DCRD-2} compared to the reference \textbf{DCRD} is attributed to the incorporation of thiophene and fluoren units as molecular bridges that enhance the injection of electrons to the electron-acceptor fragment. Furthermore, the \textbf{DCRD-2} system has a strong electron-tracting group, which will have greater absorption in the UV-Vis region, thus promoting charge transfer with greater efficiency than the compound reference \textbf{DCRD} and the compound \textbf{DCRD-1}. Therefore, $E_g^{el}$ increases in the order \textbf{DCRD}> \textbf{DCRD-1}> \textbf{DCRD} both in the gas phase and in EtOH. In other words, the modification of the molecular bridges creates a relatively smaller band gap between the orbitals and efficient CT towards the electron-acceptor fragment from the electron-donor fragment, demonstrating that the systems \textbf{DCRD-1} and \textbf{DCRD-2} are efficient materials for OSC, compared to the reference dye \textbf{DCRD}.

\subsection{Electronic Absorption Spectral}
\medskip

\begin{figure}[htp]
\centering
\includegraphics[scale=0.65]{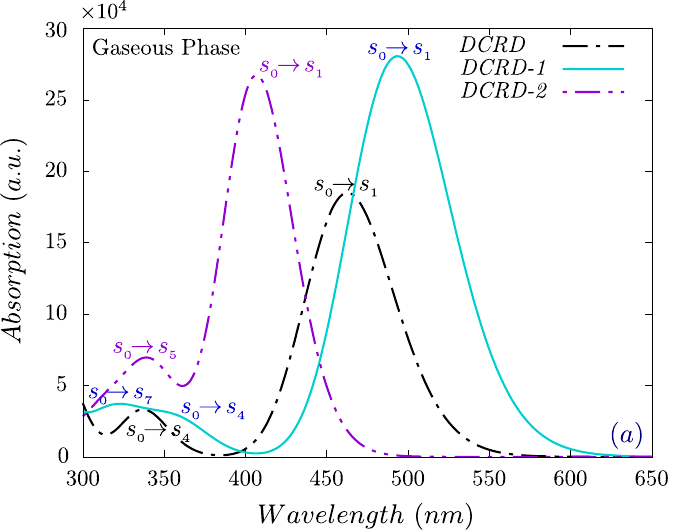}\includegraphics[scale=0.65]{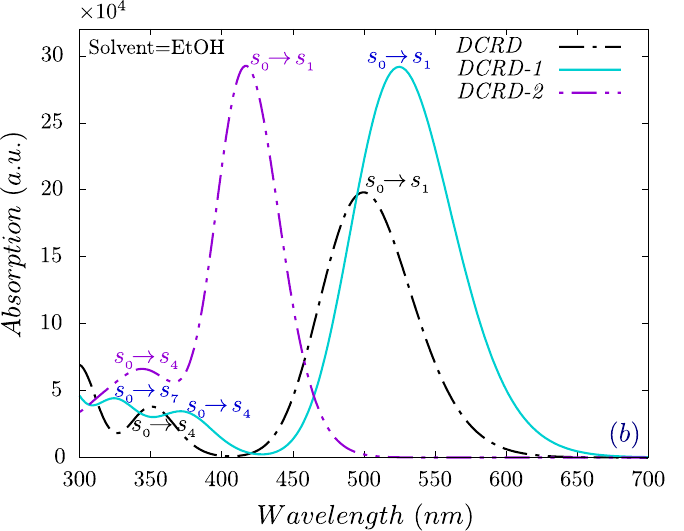}
\caption{UV-Vis absorption spectra of molecular systems based on triphenylamine and rhodanine obtained using TD-DFT with the CAM-B3LYB/6-31g(d,p) basis set in the gaseous phase and ethanol solvent.}
\label{Fig_3}
\end{figure}
To elucidate the effect of molecular bridges on the optical properties of new systems, molecules with fascinating light-harvesting capacity are useful to improve the productivity of photovoltaic devices. The electronic transition energies ($E_{xc}$), the oscillator strength ($f_{osc}$) and the maximum absorption lengths ($\lambda_{max}$) of the molecular systems were studied (\textbf {DCRD}--\textbf{DCRD-2}) in the gas phase. In the solvent phase at the TD-DFT/CAM-B3LYP/6-31+(d,p) level of theory, the $\lambda_{max}$ and other parameters are illustrated in Table~\sugb{\ref{tab1}}. Our model and the molecular systems designed in the EtOH solvent present a broad and intense absorption peak ranging between $417.69$ and $525.55$~nm, while for the gas phase it occurs between $407.60$ and $494.83$~nm as shown in Fig.~\sugb{\ref{Fig_3}}. It is also observed that the reference \textbf{DCRD} system has $\lambda _{max}=499.62$~nm in the presence of EtOH, which is in accordance with the value recorded experimentally ($489$~nm), as well as a peak absorption at $347.78$~nm, corresponding to transitions of $S_0\rightarrow S_1$, and $S_0\rightarrow S_4$, respectively. The high energy transition at $499.62$~nm (HOMO$\rightarrow$LUMO) can be assigned to an intramolecular charge transfer between the \textbf{TPA} donor unit and the \textbf{DCR} unit, while the low energy transition energy is given at $347.78$~nm (H-1$\rightarrow$L) corresponds to a charge reorganization between the molecular bridging unit \textbf{B1} and the unit \textbf{DCRD}. In the case of the system \textbf{DCRD-1}, three absorption peaks are observed, where the maximum absorption occurs at $525.55$~nm (for $S_0\rightarrow S_1$) corresponding to the HOMO$\rightarrow$LUMO electronic transition of greatest contribution, the following two peaks appear around $369.69$~nm (for $S_0\rightarrow S_4$) and $326.14$~nm (for $S_0\rightarrow S_7$), corresponding to the electronic transitions of H-3$\rightarrow$L+1. The first and second absorption bands ($525.55$~nm and $369.69$~nm) can be assigned to an intramolecular charge transfer from the 4-Hexylocy-Triphenylamine fragment to the 2-(1,1-Dicyanomethylene)-1,3-thiazolene-4 fragment, while for the third energy band, it corresponds to a charge reorganization inside the fragment composed of the molecular bridge and the acceptor unit. The compound \textbf{DCRD-2}, on the contrary, only presents two absorption peaks, the first around $417.69$~nm and the second around $347.79$~nm, corresponding to electronic transitions of HOMO$\rightarrow$LUMO and HOMO$\rightarrow$L+1, respectively. It can also be observed that the compound \textbf{DCRD-1} presents a bathochromic shift. In contrast, the molecular system \textbf{DCRD-2} presents a hypsochromic change in the absorption spectrum, compared to the spectrum of \textbf{DCRD} as shown in Fig.~\sugb{\ref{Fig_3}}, both in the EtOH solvent and in the gas phase.
\medskip

\begin{table*}[ht] 
\begin{center}
\centering
\caption{Wavelengths of the most important simulated transition states $\lambda$, oscillator strengths $f_{os}$, excitation energy $E_{ex}$, and light harvesting eﬃciency $LHE$.}
\scalebox{1.2}{
\begin{tabular}{w{c}{2cm}w{c}{1cm}w{c}{1.5cm}w{c}{1.6cm}w{c}{1.5cm}w{c}{1.2cm}m{3.5cm}} 
\hline
\toprule
&&&&&\\
System & States & $\lambda$ (nm) & $\Delta E_{ex}$ (eV) & $f_{osc}$ & LHE   \\ 
&&&&&\\
\hline
\toprule
                &          &        & \textit{\normalsize{Gas Phase}} & &     \\
\midrule        
                & $S_0\rightarrow S_{1}$  & 462.40 & 2.681 & 2.549 & 0.9972   \\ 

\textbf{DCRD~~~}   & $S_0\rightarrow S_{4}$  & 335.04 & 3.701 & 0.292 &       \\
\midrule
                & $S_0\rightarrow S_{1}$  & 494.83 & 2.506 & 3.368 &         \\ 

\textbf{DCRD-1} & $S_0\rightarrow S_{4}$  & 355.31 & 3.489 & 0.159 & 0.9995  \\ 

                & $S_0\rightarrow S_{7}$  & 318.93 & 3.888 & 0.243 &         \\ 
\midrule
                & $S_0\rightarrow S_{1}$  & 407.60 & 3.042 & 3.238 &         \\

\textbf{DCRD-2} & $S_0\rightarrow S_{5}$  & 337.73 & 3.671 & 0.569 & 0.9994  \\
\midrule
                &          &        & \textit{\normalsize{EtOH}} & &         \\
\midrule
                & $S_0\rightarrow S_{1}$  & 499.62 & 2.482 & 2.725 & 0.9981  \\ 

\textbf{DCRD~~~}   & $S_0\rightarrow S_{4}$  & 347.78 & 3.565 & 0.320 &      \\
\midrule
                & $S_0\rightarrow S_{1}$  & 525.55 & 2.359 & 3.509 &         \\ 

\textbf{DCRD-1} & $S_0\rightarrow S_{4}$  & 369.69 & 3.354 & 0.233 & 0.9996  \\ 

                & $S_0\rightarrow S_{7}$  & 326.14 & 3.802 & 0.397 &         \\ 
\midrule
                & $S_0\rightarrow S_{1}$  & 417.69 & 2.968 & 3.447 &         \\

\textbf{DCRD-2} & $S_0\rightarrow S_{4}$  & 347.79 & 3.565 & 0.404 & 0.9996  \\
\bottomrule
\hline
\end{tabular}}
\label{tab2}
\end{center}
\end{table*}

For all molecular compounds, the highest $\lambda_{max}$ of $525.55$~nm and the lowest excitation energy of $2.3591$~eV were observed in \textbf{DCRD-2} when the employed bridge molecular of 2-(9,9-Dihexyl-9H-Fluoren-2-yl)hexylthiophene. This improvement is due to the introduction of fluorene, which enhances the system's absorption capacity and charge transfer compared to systems with thiophene fragments. The combined electron-attracting nature of the halo and -CN groups in the \textbf{DCRD} system reduced the inter-orbital $E_g^{el}$ and excitation energy with a broader absorption band. The general descending order of $\lambda_{max}$ of the compounds studied both in EtOH, and in the gas phase,~\textbf{DCRD-1}>\textbf{DCRD}>\textbf{DCRD-2}.
\medskip

On the other hand, the strength of the oscillator ($f_{osc}$)~\sugb{\cite{rashid2022}} is the probability of electronic transition from the HOMO levels of the ground to the LUMO levels excited in the photovoltaic cell. Higher absorption in the UV-visible zone means higher $f_{osc}$, which shows a higher charge transfer rate and, therefore, a higher light harvesting efficiency (LHE) since LHE measures a molecule's ability to absorb photons. Furthermore, it is a vital component directly correlated with OSCs efficiency under incident light~\sugb{\cite{rani2022}}. Molecules expressing high LHE exhibit high short-circuit current ($J_{sc}$) and are more efficient. The $f_{osc}$ and LHE are related by Equation~(\sugb{\ref{ecu1}})~\sugb{\cite{rani2022}}.
\begin{equation}
LHE=1-10^{-f_{osc}}
\label{ecu1}
\end{equation}
Table~\sugb{\ref{tab1}} shows the LHE for the studied molecules, in which it is observed that \textbf{DCRD}, \textbf{DCRD-1}, and \textbf{DCRD-2} have oscillator intensities in the range of $2.549$ to $3.368$ in the gas phase and $2.725$ to $3.509$ in the EtOH solvent (Table~\sugb{\ref{tab2}}). The molecularly bridged structure \textbf{DCRD-1} that facilitates photon absorption has higher $f_{osc}$ and, as a result, higher absorption and charge transformations. Furthermore, in Table~\sugb{\ref{tab2}}, it is observed that \textbf{DCRD} has shown relatively lower oscillator strength values due to the poor overlap of the orbitals involved in the electronic transitions. Also, the similar symmetry of the orbitals contributing to the electronic transitions causes lower ~$\pi\rightarrow\pi^*$ transitions, resulting results in lower oscillator strength. \textbf{DCRD-1} is expexted to show high photocurrent among all the designed molecules because it exhibits higher oscillator strength.

\subsection{Ionization Potential and Electron Affinity.}
\medskip
Ionization potential (IP) and electron affinity (EA) are useful tools to analyze the charge transfer efficiency in solar cells~\sugb{\cite{louis2017}}. Molecules having electron donating groups reveal low IP values because they destabilize the HOMO energy level by facilitating electron transfer~\sugb{\cite{ahmed2019}}. On the contrary, those possessing electron-withdrawing groups exhibit higher IP values due to the stabilization of the HOMO level, and it is challenging to remove electrons~\sugb{\cite{ahmed2019}}. The IP and EA of all the molecules considered can be determined from Equation~\sugb{\cite{waqas2022}},
\begin{eqnarray}
&IP& = \left( E_{0}^{+} - E_{0} \right),\nonumber \\
&EA& = \left( E_{0} - E_{0}^{-} \right),
\label{ecu2}
\end{eqnarray}

Here, the neutral molecule has a cation characterized by energy $E_{0}^{+}$ and an anion characterized by energy $E_{0}^{-}$. The energy of a neutral molecule is denoted by the symbol $E_0$~\sugb{\cite{amati2019, pereira2023}}. Table~\sugb{\ref{tab3}} displays the calculated IP and EA values for the studied compounds. The optimized \textbf{DCRD-1} and \textbf{DCRD-2} molecules were found to have higher charge transfer efficiency than \textbf{DCRD}, which was determined by the findings that the IP values for the optimized molecules were lower than those of \textbf{DCRD}. \textbf{DCRD-2} has the lowest IP ($5.198$~eV) and the highest charge extraction from the HOMO. The IP value of the \textbf{DCRD} molecule is high, indicating poor charge transfer. Each synthesized molecule, \textbf{DCRD-1} and \textbf{DCRD-2}, has higher electron affinity than the standard reference \textbf{DCRD}.

\subsection{Reorganization Energy.}
\medskip
On a microscopic scale, two key parameters determine charge transport efficiency in organic systems: the electronic coupling ($V_{ij}$) between electron-acceptor and electron-donor fragments and the reorganization energy ($\lambda$), which influences the operational efficiency of solar cells. Maximizing $V_{ij}$ and minimizing $\lambda$ increases charge carrier mobility. Reorganization energy generally divides into internal reorganization energy ($\lambda_{int}$) or intramolecular reorganization energy and external reorganization energy ($\lambda_{ext}$)~\sugb{\cite{li2015}}. External deformations in the molecule relate to $\lambda_{ext}$, which is excluded due to the external solvent environment. In contrast, $\lambda_{int}$ is related to changes in the internal environment, such as variations in geometry/structure~\sugb{\cite{akram2022}}. The intramolecular reorganization energy $\lambda_{int}$ can be estimated for electrons (reorganization energy of electrons $\lambda_{e}$) and for holes (reorganization energy of holes $\lambda_{h}$) and can be expressed by the following equation:~\sugb{\cite{rafiq2022}}
\begin{eqnarray}
&\lambda_{e}& = \left( E_{0}^{-} - E_{-} \right) + \left( E_{-}^{0} - E_{0} \right)\\
\nonumber\\
&\lambda_{h}& = \left( E_{0}^{+} - E_{+} \right) + \left( E_{+}^{0} - E_{0} \right),
\label{ecu3}
\end{eqnarray}

where $E_0^-$ and $E_0^+$ are anionic and cationic energies used in these equations to describe the ground state energy of a neutral compound. At a single-point ground state, $E_0$ represents the energy of a neutral molecule, whereas $E_-$ and $E_+$ reflect the anion and cation single-point ground states energies after optimization. $E_-^0$ and $E_+^0$ are the ground-level energies for anions and cations after optimization, respectively.
\medskip

\begin{table}[htp] 
\begin{center}
\centering
\caption{Estimated values for the electron reorganization energy ($\lambda_e$) along with the hole reorganization energy ($\lambda_h$), ionization potential (IP), electron affinity (EA), and Dipole Moment ($\mu$) of all Researched Molecular dyes in EtOH and phase gas.}
\scalebox{0.92}{
\begin{tabular}{w{c}{1.5cm}w{c}{1.5cm}w{c}{1.3cm}w{c}{1.3cm}w{c}{1.3cm}w{c}{1.3cm}} 
\hline
\toprule
\normalsize{System} & $\lambda_{e}$ (eV) & $\lambda_{h}$ (eV) & $IP$ (eV) & $EA$ (eV) & $\mu$ (D)\\ 
\midrule
                 & &\textit{\normalsize{EtOH}} & & &  \\ 
\textbf{DCRD~~~} & 0.316 & 0.201 & 5.379 & 3.284 & 13.504 \\ 
\textbf{DCRD-1}  & 0.142 & 0.233 & 5.302 & 3.325 & 16.762 \\
\textbf{DCRD-2}  & 0.317 & 0.198 & 5.198 & 3.374 & 18.258 \\ 
\midrule
                 & &\textit{\normalsize{Gas Phase}} & & &  \\ 
\textbf{DCRD~~~} & 0.261 & 0.220 & 6.768 & 2.084 & ~9.063 \\ 
\textbf{DCRD-1}  & 0.164 & 0.285 & 6.502 & 2.121 & 11.672 \\ 
\textbf{DCRD-2}  & 0.219 & 0.209 & 6.309 & 1.937 & 13.042 \\
\bottomrule
\hline
\end{tabular}}
\label{tab3}
\end{center}
\end{table}

The results for the reorganization energy of both electrons ($\lambda_e$) and holes ($\lambda_h$) are shown in Table~\sugb{\ref{tab3}}. Understanding the relationship between a material molecular structure and charge transport properties is critical in designing promising candidates for solar cell devices. Furthermore, it is well known that the lower the values of $\lambda$, the higher the charge transport rate~\sugb{\cite{sun2017}}. The data in Table \sugb{\ref{tab3}} shows that the values of $\lambda_h$ for \textbf{DCRD} and \textbf{DCRD-2} are all smaller than their respective $\lambda_e$; this suggests that the hole transfer rate is greater than the electron transfer rate. The hole reorganization energies $\lambda_h$ calculated for the systems \textbf{DCRD} and \textbf{DCRD-2} are lower than those of N,N'-diphenyl-N,N'-bis(3-methylphenyl)-(1,1'-biphenyl)-4,4'-diamine (TPD) which is a typical hole transport material $\lambda_h=0.290$~eV~\sugb{\cite{gruhn2002}}. This implies that the hole transfer rates of \textbf{DCRD} and \textbf{DCRD-2} could be higher than those of TPD. Therefore, \textbf{DCRD} and \textbf{DCRD-2} could be hole transport materials (HTM) from the point of view of the lowest reorganization energy. Furthermore, it can be observed that when comparing the reorganization energy both in the gas phase and in the EtOH solvent, it is found that the $\lambda$ are lower in EtOH, which implies that EtOH favors the transport processes. On the other hand, the $\lambda_e$ for the compound \textbf{DCRD-1} is lower than the $\lambda_h$, so the electron transfer rate will be greater than the hole transfer rate. Furthermore, the value of $\lambda_e$ of the molecular system \textbf{DCRD-1} is smaller than that of tris(8-hydroxyquinolinato)aluminum(III) (Alq3), which is a typical electron transport material $\lambda_e=0.276$~eV~\sugb{\cite{irfan2009}}. This indicates that its electron transfer rate could be higher than that of Alq3. Therefore, \textbf{DCRD-1} can be used as promising electron transport materials in organic light-emitting diodes (OLEDs) from the point of view of the minor rearrangement energy. Likewise, electronic transfer processes are more favorable in the presence of EtOH than in the gas phase. These results indicate that the values of $\lambda_h$ and $\lambda_e$ are affected by introducing different molecular bridges and substitution groups in these molecules. The introduction of a molecular bridge of 2,5--Dimethylthiophene and 2--(9,9--Dihexyl--9H-Fluoren--2--yl)hexylthiophene leads to an increase in hole transfer rates, while the molecular bridge 4,4''--Dihexyl--2,2':5',2''--Terthiophene decrease hole transfer rates, favoring electronic transfer processes. On the other hand, the differences between the values $\lambda_e$ and $\lambda_h$ are lower for the molecular system \textbf{DCRD-1}. It implies that \textbf{DCRD-1} have better equilibrium properties for hole and electron transport. Therefore, it can be used as a good candidate for ambipolar charge transport materials under appropriate operating conditions.

\subsection{Molecular electrostatic potential}
\medskip
To better explain how intramolecular charge distribution occurs three-dimensionally and explain electron-rich and electron-poor sites, molecular electrostatic potential (MEP) measurements were performed computationally. The MEP has proven to be effective in evaluating the degree of charge transfer between the charge donor and acceptor fragments~\sugb{\cite{zhang2022}}. The MEP is the three-dimensional visualization of the charge density at different positions of a molecule of interest. The MEP can predict the presence of electrons, lone pairs, and electron-withdrawing species on the surface of molecules. These MEP plots show several colors, each representing an additional molecule feature. Here, the red color indicates the negative value of the electrostatic potential and an abundance of electrons in that region, which is undoubtedly located above the acceptor region where electronegative atoms may be present; the green color shows an electrically neutral region. In contrast, the blue color illustrates the positive value of the electrostatic potential accompanied by the absence of electrons found mainly on the electron-donor fragment (central nucleus) and adjusts to the nucleophilic reactivity~\sugb{\cite{hassan2022, hussain2023}}.

\begin{figure}[htp]
\centering
\includegraphics[scale=0.15]{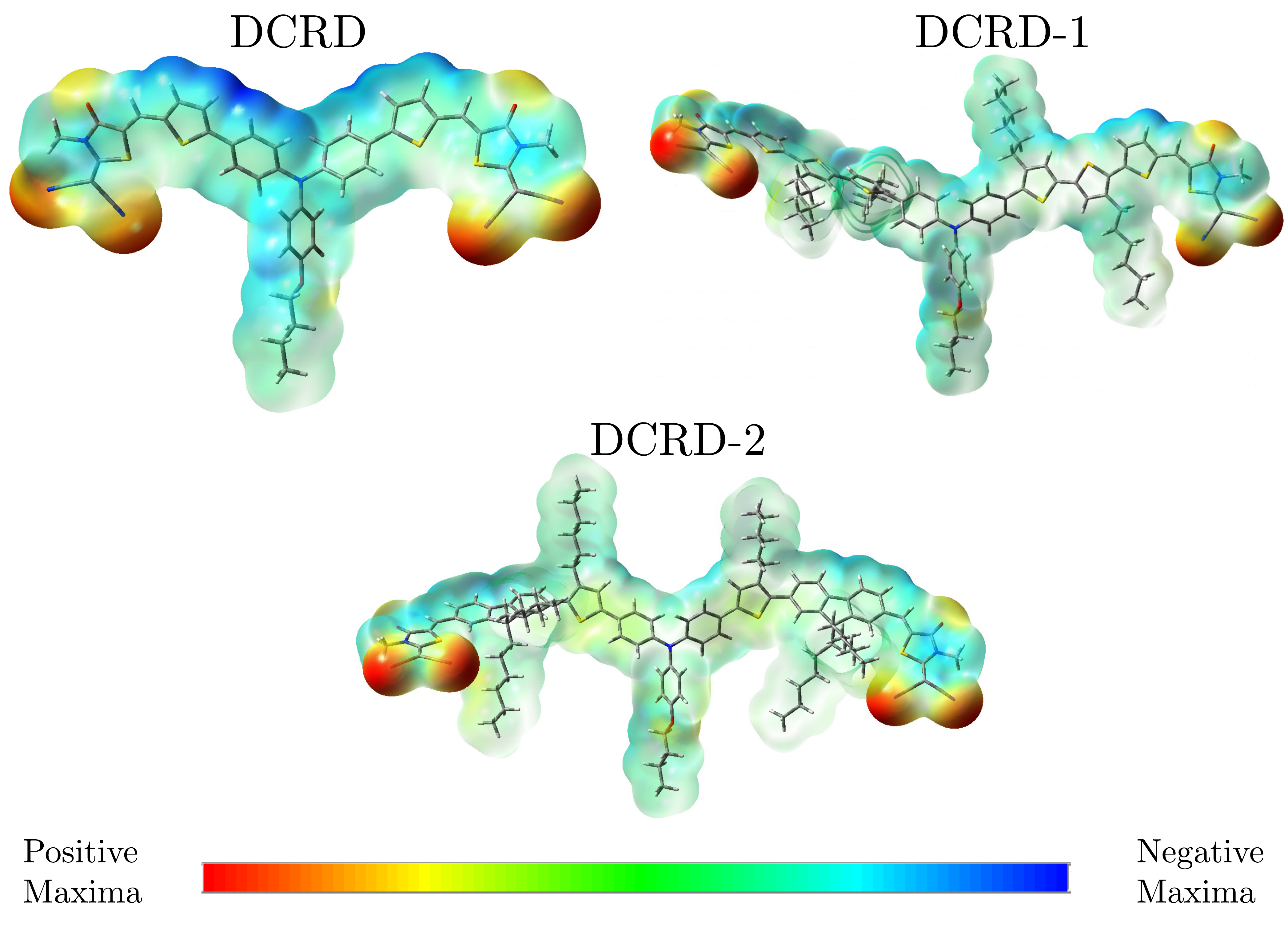}
\caption{Molecular electrostatic potential (MEP) maps of the molecular compounds \textbf{DCRD}, \textbf{DCRD-1} and \textbf{DCRD-2}.}
\label{Fig_4A}
\end{figure}

Fig.~\sugb{\ref{Fig_4A}} illustrates that the $\pi$-bridge and the central core have fewer electrons than the terminal parts revealed by the green color (best seen in \textbf{DCRD-1} and \textbf{DCRD-2}). The red terminal acceptor groups indicate a higher negative charge due to unsaturated electronegative atoms: oxygen, nitrogen, and sulfur, suggesting a high electron density appropriate for electrophilic attacks. The compound \textbf{DCRD} presents a bluer color on the acceptor and the triphenylamine core of the donor part, which reveals that these parts have a more positive IP. Because the electronegative atoms are located at a peripheral site, the proposed compound \textbf{DCRD-1} showed a redder color (negative charge) than any of the others, demonstrating that this molecule has a higher negative IP. The presence of different regions based on electron density implies that the designed molecules are ideal candidates for efficient OSCs.

\section{Photovoltaic Properties}\label{photovoltaic}

For OPV technology, the most efficient cells are bulk heterojunction (BHJ) devices, as BHJ-based organic solar cells (OSCs) have a promising future to convert solar energy into electricity more economically than conventional solar cells of silicon~\sugb{\cite{gurney2019, ma2019}}, due to their low-temperature manufacturing process, light weight and flexibility~\sugb{\cite{scharber2013, yu2014}}, and its conversion efficiency energy reaching more than 19\%~\sugb{\cite{zhu2022}}. Although the PCE of OPV cells has been improved to some extent, there is still much work to be done to achieve their commercial application compared to the 25.2\% PCE of perovskite-based thin film solar cells~\sugb{\cite{liang2023}}. There is a need to continue improving the performance of the IP. The power conversion efficiency, PCE, depends directly on the open circuit voltage ($V_{oc}$), the density of short-circuit current ($J_{sc}$), the fill factor ($FF$) and the energy solar incident. None of these four components can be ignored in material design to achieve better PCE. Generally, Voc is related to the energy difference between the HOMO of the donor material and the LUMO of the acceptor material. It is also affected by carrier generation and recombination rates as well as tail energy states or trap states~\sugb{\cite{sweetnam2014}}.

\subsection{Open Circuit Voltage ($V_{oc}$) Investigation.}
\medskip
The open circuit voltage ($V_{OC}$) significantly contributes to determining the photovoltaic characteristics of organic photovoltaic devices, since the performance of OSCs is highlighted by the open circuit voltage (Voc)~\sugb{\cite{tang2021}}. It accounts for the most significant amount of current of any optical material. The highest voltage drawn from a photovoltaic device occurs at zero current level~\sugb{\cite{wang2019}}. The $V_{oc}$ is influenced by several factors, such as device temperature, energy levels of the molecules, external fluorescence, light source, and charge carrier recombination~\sugb{\cite{azzouzi2019}}. The open circuit voltage is directly associated with the energy difference of HOMO and LUMO of the Acceptor and Donor molecules, i.e., $E_{_H}^D$--$E_{_L}^A$. In the current research, a renowned acceptor, PC$_{61}$BM, is used, which allows the design of three BHJ OSCs with active mixing layer: (1) \textbf{DCRD}:\textbf{PC$_{61) }$BM}, (2) \textbf{DCRD-1}:\textbf{PC$_{61}$BM}, (3)~\textbf{DCRD-2}:\textbf{PC$_{61} $BM}. Energy offsets at D-A interfaces within the active layer of an OSC affect both $V_{oc}$ and $J_{sc}$ and, therefore, the PCE~\sugb{\cite{afzal2020, hussain2020}}. Theoretically, the examined CSO Voc results are determined using the equation~\sugb{(\ref{ecu4})}, reported by Scharber et al~\sugb{\cite{scharber2013}}, and the results are shown in Table~\sugb{\ref{tab3}}.
\begin{equation}
V_{oc}=\frac{1}{e}\Big\vert E_{_{LUMO}}^A- E_{_{HOMO}}^D\Big\vert-0.3~V
\label{ecu4}
\end{equation}

Where e is the elementary charge, the value of $0.3$~V is a typical loss in bulk heterojunction solar cells. And using $E_L=-3.68~eV$ simulated in this work for of PC$_{61}$BM lowest occupied molecular orbital (LUMO) energy.

\begin{figure}[htp]
\centering
\includegraphics[scale=0.3]{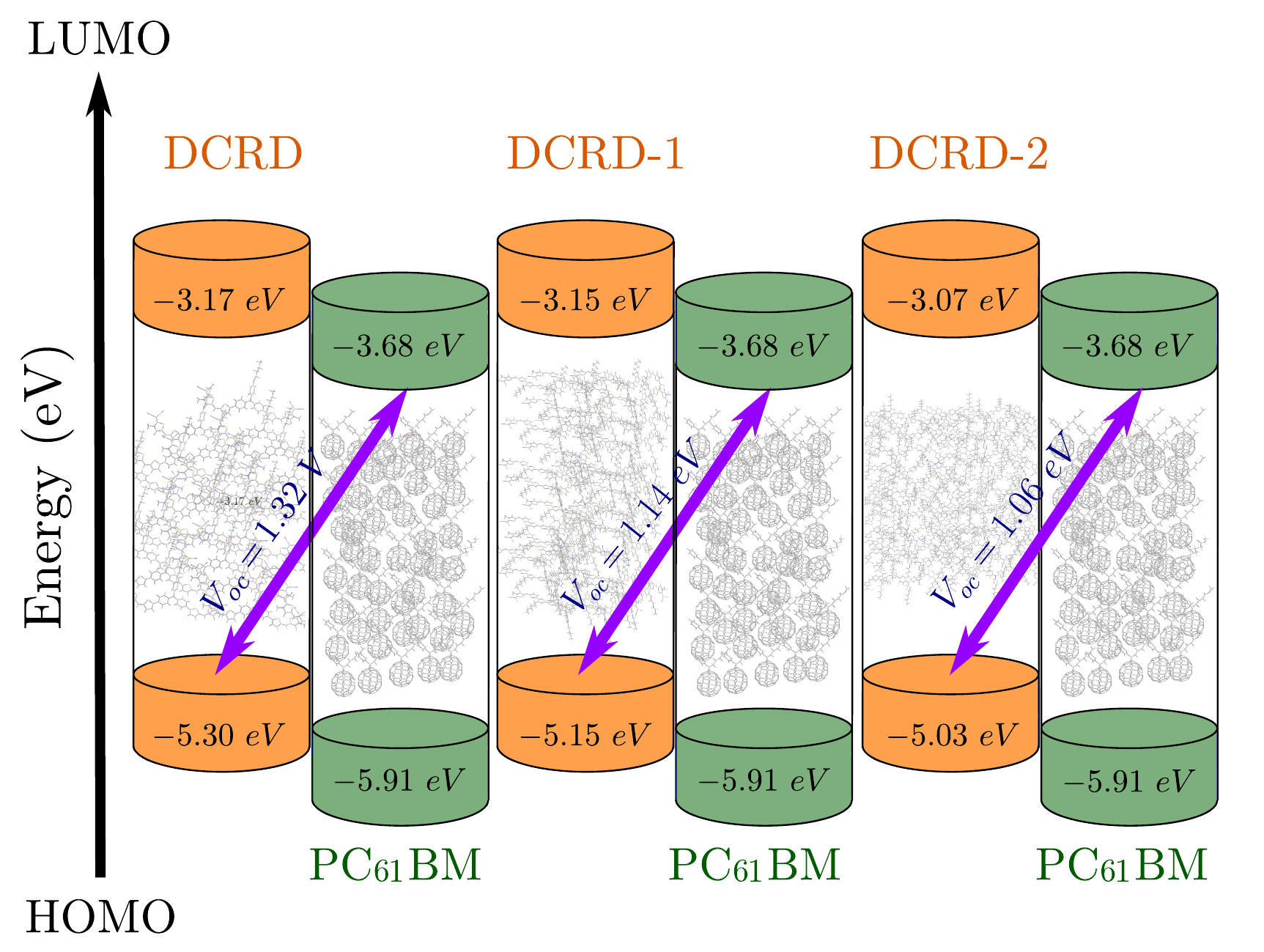}
\caption{Variation of $V_{OC}$ in all developed compounds using acceptor PC$_{61}$BM}
\label{Fig_4}
\end{figure}

At the same time, however, for efficient dissociation of charge transfer (CT) excitons formed at D/A interfaces, the energy difference in HOMO and LUMO of the donor molecule and the acceptor molecule at the interface, $\Delta E=E_{_{LUMO}}^D-E_{_{LUMO}}^A$, must be greater than or equal to the binding energy of the CT excitons ($\Delta E_{_{LUMO}} \geq E_B$)~\sugb{\cite{setsoafia2023}}.

In operational solar cells, additional recombination paths decrease the value of $V_{oc}$ and affect the PCE of the OPVs because they must operate at a voltage lower than $V_{oc}$. Based on these considerations, the ideal parameters for a $\pi$-conjugated-system:PC$_{61}$BM device are shown in Table~\sugb{\ref{tab3}}. The $V_{oc}$ values of the $DCRDs$ systems are in the range of $1.06-1.32$~V, indicating values greater than $1.0$~V efficiently increasing the value of $V_{oc}$; furthermore, since $\Delta E$ shows values higher than $0.3~$eV, this suggests that the transfer of photoexcited electrons is accessible from the hole transport material to the acceptor PC$_{61}$BM.
\begin{table}[htp] 
\begin{center}
\centering
\caption{Parameter values of the studied molecular compounds: Energy of the exciton
driving force $\Delta E$, open circuit voltage ($V_{oc}$), short-circuit current density ($J_{sc}$), fill factor ($FF$), and power conversion efficiency (PCE) in the solvent EtOH.}
\scalebox{0.91}{
\begin{tabular}{w{c}{1.3cm}w{c}{1.3cm}w{c}{1.3cm}w{c}{2cm}w{c}{1.3cm}w{c}{1.3cm}} 
\hline
\toprule\\
\normalsize{System} & $\Delta E$ (eV) & $V_{oc}$ (V) & $J_{sc}$ (mA/cm$^2$) & $FF$ & $\eta$ (\%) \\ 
\midrule
\textbf{DCRD~~~} & 0.51 & 1.32 & ~9.52 & 0.810 & 10.18  \\ 
\textbf{DCRD-1}  & 0.53 & 1.14 & 12.21 & 0.786 & 10.95  \\
\textbf{DCRD-2}  & 0.63 & 1.06 & 11.85 & 0.774 & ~9.72   \\ 
\bottomrule
\hline
\end{tabular}}
\label{tab4}
\end{center}
\end{table}
As stated above, the value of $V_{oc}$ depends mainly on the level of $E_{_{HOMO}}^D$ and $E_{_{LUMO}}^A$. A lower acceptor origin LUMO results in a higher $V_{oc}$ value and improved optoelectronic parameters. The HOMO electron movement of donor molecules is increased by low acceptor LUMO, which directly improves the optoelectronic characteristics. Furthermore, the HOMO-LUMO bandgap between the donor and acceptor units amplifies the PCE values. As a result, Fig.~\sugb{\ref{Fig_4}} represents the orbital energy diagram of the chromophores mentioned above relative to PC$_{61}$BM. Fig.~\sugb{\ref{Fig_4}} shows that the LUMO levels of the donor molecular compounds \textbf{DCRD}, \textbf{DCRD-1} and \textbf{DCRD-2} are higher than the level of LUMO of the acceptor chromophore. This facilitates the transfer of electrons from the donor polymers to the acceptor segment, which enhances the molecular optoelectronic parameters studied. Furthermore, among the structures developed, it has been shown that \textbf{DCRD} ($1.32$~V) and \textbf{DCRD-1} ($1.14$~V) are the molecules with the highest $V_{OC}$ due to the participation of the thiophene entity in the molecular bridges \textbf{B1} and \textbf{B2}, from which it is concluded that \textbf{DCRD} and \textbf{DCRD-1} exhibit more charge delocalization, conjugation and, therefore, Therefore, they will have higher PCE than the \textbf{DCRD-2} molecule. It is found that the Voc results of the molecular compounds are in the decreasing order of \textbf{DCRD} > \textbf{DCRD-1} > \textbf{DCRD-2}.

\subsection{Fill Factor (FF)}
\medskip
The PCE and other photovoltaic properties of OSCs can also be estimated by utilizing a compelling tool, the Fill Factor (FF). In practice, many physical mechanisms contribute to this, and consequently, many models have been suggested to estimate the FF~\sugb{\cite{jao2016,qi2013}}. Generally, the suggested models are based on the relationship between current and voltage; but with different assumptions about the causes and values of shunt and series resistances. The FF is usually represented as a function of $V_{oc}$. The filling factor of molecular compounds can be measured with the application of Equation (\sugb{\ref{ecu5}})~\sugb{\cite{alharbi2015}}:
\begin{equation}
FF=\frac{V_{oc}}{V_{oc} + \alpha k_BT}
\label{ecu5}
\end{equation}
where $\alpha$ is a factor that can be adjusted for other solar cell technologies, for example, $\alpha=6$ and $\alpha=12$ best fit the upper limits of the FF measured for excitonic solar cells and non-excitonic, respectively, as shown by Alharbi et al.~\sugb{\cite{alharbi2015}}. In this equation, $V_{oc}$ represents open circuit voltage, T characterizes the temperature of the system that was $300$~K, Boltzmann constant is symbolized by $k_B$. The calculated fill factor values for the reference molecule and the designed chromophores (\textbf{DCRD-1} and \textbf{DCRD-2}) are summarized in Table~\sugb{\ref{tab4}}, which are 0.810, 0.786, and 0.774, respectively. The measured results displayed that the molecules \textbf{DCRD}, and \textbf{DCRD-1} have almost similar fill factor values, which is why they exhibit admirable photovoltaic properties compared with \textbf{DCRD-2} having lower FF value. All compounds descending pattern of FF follows the sequence: \textbf{DCRD} > \textbf{DCRD-1} > \textbf{DCRD-2}. Higher FF values of \textbf{DCRD} and \textbf{DCRD-1} caused an increase in PCE compared to the \textbf{DCRD-2} system. Furthermore, it should be noted that the increasing trend of FF is similar to that of $V_{oc}$, and their results showed that the designed composites are of better quality than \textbf{DCRD-2} with remarkably increasing PCE.

\subsection{Short-Circuit Current Density ($J_{sc}$)}
\medskip
Current-voltage density characteristics ($J$--$V$) are necessary for the context of voltage loss analyses, which are based on values of the open circuit voltage ($V_{oc}$) and the short circuit current density ($J_{sc}$). The most accurate way to determine $J_{sc}$ is to consider the spectral shape of sunlight since the intensity and spectral shape depend on various circumstances, as well as the spectral responses of the calibration cell, the OSC medium, the angle above the horizon, by which a standard power spectrum $\phi _{AM.1.5}(\lambda)$~\sugb{\cite{American, green2020}} is defined. Therefore, $J_{sc}$ can be determined by the following expression~\sugb{\cite{el2015,madrid2021,jungbluth2022}},
\begin{equation}
J_{sc}=\frac{q}{hc}\int_{0}^{E_g^D}EQE(\lambda)P_{AM1.5G}(\lambda) \lambda d\lambda
\label{Ecu7}
\end{equation}
where $h$ is Plank’s constant, $c$ is the speed of light in vacuum, $E_g^D$ is the band gap of the hole transporting material in the active layer of the device, and $EQE(\lambda)$ is the external quantum efficiency, which for this work is assumed with a value of $80\%$~\sugb{\cite{scharber2013,li2015}}.
The values of $J_{sc}$ and PCE for the systems \textbf{DCRD}-\textbf{DCRD-2} are shown in Table~\sugb{\ref{tab4}}, in which it is observed that the values of $J_{sc}$ are in a range of $9.52$--$12.21$~mAcm$^2$, with the highest value found for the molecular system \textbf{DCRD-1}, which also has the highest value of PCE around 10.95\%. This study shows, concerning the PCE of the systems considered and taking the system \textbf{DCRD} as a reference, that the inclusion of thiophene rings as molecular bridges to generate the system \textbf{DCRD-1}, favors the increase of the PCE. However, it is observed that by including dibenzothiophene (\textbf{DCRD-2}) as a molecular bridge, a significant PCE reduction of around 1.23\% is obtained; this shows that both the molecular system \textbf{DCRD} and \textbf{DCRD-1}, the latter showing better results, present good results for use as photovoltaic materials in BHJ organic solar cells compared to \textbf{DCRD-2}.
\section{Conclusions}\label{Conclusions}
We have studied three molecular systems, A-$\pi$-D-$\pi$-A, that contain triphenylamine and 2-(1,1-dicyanomethylene)rhodanine in their structure. The geometric, electronic, optical and photovoltaic properties of the designed systems were investigated using DFT calculations with the B3LYP/6-31G(d,p) and TD-DFT/CAM-B3LYP/6-31G(d,p) levels of theory. The energy estimation of the HOMO-LUMO limiting molecular orbitals correlated significantly with the the available experimental data~\sugb{\cite{echeverry2014}}. The exciton driving force energy ($\Delta E$) of all molecular systems (\textbf{DCRD-DCRD-2}) has a value greater than $0.3$~eV, which can ensure efficient exciton splitting. The simulated UV/VIS absorption spectra show a similar profile for all systems, presenting a robust main band between $525.55$~nm and $417.69$~nm. In general, the \textbf{DCRD} and \textbf{DCRD-1} molecular systems have the best photovoltaic properties compared to the \textbf{DCRD-2} derivative, where low values in the gap energy ($E_g^{el}$) were observed ($2.13-1.97$~eV for \textbf{DCRD} and \textbf{DCRD-1}, respectively) and higher value in short-circuit photo-current density ($J_{sc}$) for \textbf{DCRD-1}, as well as better photoelectric conversion efficiency (PCE) (10.95\%), so these derivatives can potentially be used in volumetric heterojunction (BHJ) organic solar cells (OSCs).
\section*{Acknowledgment}
The authors acknowledge the University of Sucre, Colombia, for financial support during the conduct of this study.
\printcredits
\bibliographystyle{model1-num-names}
\bibliography{cas-refs}
\end{document}